\documentclass[twocolumn,showpacs,pre,aps,superscriptaddress]{revtex4}
\usepackage[utf8]{inputenc}
\usepackage[english]{babel} 
\usepackage{graphicx} 
\usepackage{listings} 
\usepackage{amssymb} 
\usepackage{amsmath} 
\usepackage{longtable}
\usepackage{multirow} 
\usepackage{booktabs}
\usepackage{pst-all} 
\usepackage{pstricks-add}
\usepackage{units} 
\usepackage{listings} 
\usepackage{xcolor}
\usepackage{subfigure}
\usepackage{hyperref}

\usepackage{natbib}

\begin{document} 

\title{Transport in tight-binding bond percolation models}

\author{Daniel Schmidtke}

\email{danischm@uos.de}

\affiliation{Fachbereich Physik, Universit\"at Osnabr\"uck,
             Barbarastrasse 7, D-49069 Osnabr\"uck, Germany}

\author{Abdellah Khodja}

\email{akhodja@uos.de}

\affiliation{Fachbereich Physik, Universit\"at Osnabr\"uck,
             Barbarastrasse 7, D-49069 Osnabr\"uck, Germany}

\author{Jochen Gemmer}

\email{jgemmer@uos.de}

\affiliation{Fachbereich Physik, Universit\"at Osnabr\"uck,
             Barbarastrasse 7, D-49069 Osnabr\"uck, Germany}

\begin{abstract}

Most of the investigations to date on tight-binding, quantum percolation models focused on the quantum percolation threshold, i.e., the analogue to 
the Anderson  transition. It appears to occur if roughly 30\% of the hopping terms are actually present. Thus, models in the delocalized regime may 
still be substantially disordered, hence analyzing their transport properties is a nontrivial task which we pursue in the paper at hand. Using a 
method based on quantum typicality to numerically perform linear response theory we find that conductivity and mean free paths are in good accord
with results from very simple heuristic considerations. Furthermore we find that depending on the percentage of actually present hopping terms, the transport
properties may or may not be described by a Drude model. An investigation of the Einstein relation is also presented.

\end{abstract}

\pacs{
05.60.Gg, 
72.80.Ng,  
66.30.Ma,  
}

\maketitle

\section{introduction}

Percolation theory  is a well known method to describe transport properties of crystals or other systems  which feature  regular 
lattices with substantial amounts  
of defects, impurities etc. It has been vastly studied from a classical point of view \cite{Shapiro1983,Kirkpat1973}. Here usually bonds 
(or sites) are filled at random on the above lattice with a probability $p$. It turns out that, depending on the type of lattice, there 
exists some $p$ at which the probability of getting a connected cluster of bonds (sites) which extends through the whole lattice changes 
abruptly from approximately zero to approximately one. This $p$ is called the critical probability $p_c$. 

Due to increasing interest in microscopic structures, which may be significantly affected by quantum effects, percolation
models based on quantum mechanics, have also received considerable attention. Also genuinely quantum phenomena like, e.g., 
quantum hall effect 
\cite{DungLee1994}, Fermi-Bose mixtures \cite{SanKanSanch2004} or general (anti)ferromagnetic systems \cite{YuRosHaas2005}, have been
addressed by means of percolation theory.\\
The main focus of the literature on quantum percolation appears to be on the transition from the "non-transport" to the transport regime
which 
is essentially of the same type as the the well-known Anderson transition \cite{KanOht1999,Ander1957}. Much effort is dedicated to the 
determination of the quantum percolation threshold $p_q$. It is found that the 
quantum percolation threshold is greater than the classical one \citep{SouLiGr1991,KanOht1999,SchuWeiFeh2005}. For example the classical 
threshold for bond
percolation in three dimensions on a simple cubic lattice  is determined to be $p^b_c \approx 0.25$ \cite{Kirkpat1973,LorZiff1998} 
whereas the threshold for
quantum percolation has been determined to lie at $p_q \approx 0.31$ \cite{SouLiGr1991,BerkAvi1996}. However,  quantitative investigations 
of transport in the delocalized regime 
appear to be restricted to preliminary studies  close to the quantum percolation threshold \cite{SouLiGr1991,BerkAvi1996,KanOht1999}.
With the work at hand we aim in contrast at a more detailed understanding of transport in systems  whose structures
are on one hand far away from clean crystals but on the other also far from being collections of disconnected clusters. More specifically, 
we quantitatively address transport properties at bond probabilites $p$ at which almost all energy eigenstates are delocalized. Generally we expect 
(and find) diffusive behavior. Unlike the diffusive behavior in periodic systems which is restricted to finite time (and length) scales 
\cite{KuboStMeII,Gaspard1996,Steinigeweg2009b} the diffusive behavior here persists due to broken translational symmetry.

The primary motivation for the work at hand is of principal and theoretical nature. Just like it has been recently done for the 
Anderson model \cite{Steinigeweg},
 we 
intend to 
demonstrate that also in percolation models regular diffusive behavior must not necessarily be induced by decoherence sources like phonon-coupling, 
etc., but may emerge in a fully coherent set up from the electronic model itself. Furthermore, even in the absence of decoherence this transport 
behavior will be demonstrated to be in good accord with simple statistical descriptions like the Drude-model. However the considerations are 
not entirely detached from concrete experimental research. In the context of research based on ultra-cold atoms the dynamics of a 
moderate number of atoms (playing the role of electrons) subject to a trap and an underlying optical lattice (but completely isolated  
from any environment otherwise) is observed. Among the central questions are the transport properties of such coherent
systems. 
\cite{Schn2012,Hckm2010}
The 
percolation models we address below may also possibly be implemented within such an an experimental framework through a modification of the optical 
lattice. In this context bond percolation may be more convenient to implement then site percolation. Furthermore, in the context of real materials, 
percolation models may be very rough descriptions of binary mixed crystal alloys in which 
the on-site potential of one 
species exceeds the band with of a regular crystal formed by the other species. In this case the lattice would separate in two sub-lattices, 
each formed by sites occupied by the the same species only. This would correspond to site percolation rather than bond percolation, however transport
properties may be expected to behave similarly. Recent investigations on magnesium alloys indicate an massive increase of resistivity caused by the 
substitution of only a few percent of the sites by another species, this being in accord with the findings in the paper
at hand
\cite{Pan}. This is discussed in more detail in Sec. \ref{sec-semi}

The paper at hand is organized as follows. In Sec. \ref{sec-models} we introduce our one-particle, tight-binding percolation model and comment 
rather briefly on 
localization using the density 
of states and the inverse participation number in Sec. \ref{sec-loca}. Thereafter (Sec. \ref{sec-current}) we specify the  quantities of interest,
namely 
the dc-conductivity and 
current auto-correlation function as connected by linear response theory. We address those quantities numerically and employ a method 
based on ``quantum typicality'' whenever samples are required that are too large to be assessed by means of exact diagonalization.
We find hints  of a transition from "non- Boltzmann" to Boltzmann-type transport with increasing $p$. Section \ref{sec-einrel} 
establishes the validity of the Einstein relation and, based on the latter introduces a mean free path. A numerical investigation of 
this mean free path confirms the above "non- Boltzmann" to Boltzmann-transport transition. Section \ref{sec-semi} is dedicated
to a comparison of our results to experimental data on binary magnesium alloys. The paper closes with  summary and 
conclusions in Sec. \ref{sec-concl}.

\section{Tight-binding bond percolation model}
\label{sec-models}

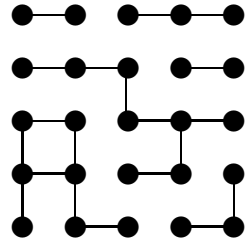
\begin{figure}
  \begin{picture}(105,100)
   \put(35.0,75.0){\circle*{8.0}}
   \put(35.0,55.0){\circle*{8.0}}
   \put(35.0,95.0){\circle*{8.0}}
   \put(35.0,35.0){\circle*{8.0}}  
   \put(35.0,15.0){\circle*{8.0}} 
   
   \put(15.0,75.0){\circle*{8.0}}
   \put(15.0,55.0){\circle*{8.0}}
   \put(15.0,95.0){\circle*{8.0}}
   \put(15.0,35.0){\circle*{8.0}}
   \put(15.0,15.0){\circle*{8.0}}
   
   \put(-5.0,75.0){\circle*{8.0}}
   \put(-5.0,55.0){\circle*{8.0}}
   \put(-5.0,95.0){\circle*{8.0}}
   \put(-5.0,35.0){\circle*{8.0}}
   \put(-5.0,15.0){\circle*{8.0}}
   
   \put(55.0,75.0){\circle*{8.0}}
   \put(55.0,55.0){\circle*{8.0}}
   \put(55.0,95.0){\circle*{8.0}}
   \put(55.0,35.0){\circle*{8.0}}
   \put(55.0,15.0){\circle*{8.0}}
   
   \put(75.0,75.0){\circle*{8.0}}
   \put(75.0,55.0){\circle*{8.0}}
   \put(75.0,95.0){\circle*{8.0}}
   \put(75.0,35.0){\circle*{8.0}}
   \put(75.0,15.0){\circle*{8.0}}
   
   

    \put(34.0,55.0){\line(0,1){19}}
    \put(55.0,75){\line(1,0){19}}
    \put(15.0,75){\line(1,0){19}}
   \put(-2.0,75){\line(1,0){19}}
   \put(15.0,15){\line(1,0){19}}
   \put(55.0,15){\line(1,0){19}}
   \put(55.0,55.2){\line(1,0){19}}
    \put(35.0,55.2){\line(1,0){19}}
   \put(15.0,15){\line(0,1){19}}
   \put(55.0,35){\line(0,1){19}}
   \put(75.0,15){\line(0,1){19}}
   
   \put(-2.0,55){\line(1,0){19}}
   \put(-2.0,35){\line(1,0){19}}
   \put(15.0,35){\line(0,1){19}}
   \put(-5.0,35){\line(0,1){19}}
   \put(-5.0,15){\line(0,1){19}}
   
   \put(38.0,35){\line(1,0){19}}
   
   \put(38.0,95){\line(1,0){19}}
   \put(58.0,95){\line(1,0){19}}
   
   \put(-2.0,95){\line(1,0){20}}
   
      
  \end{picture}
  
  \caption{Two dimensional model of classical bond percolation with $p=0.5$. In classical considerations this describes the situation 
  right at the percolation threshold, whereas from a quantum point of view this model would be below the quantum percolation threshold, 
  even in a three dimensional version. This is due to an effect comparable to Anderson localization.}
  \label{pic:modperc}
  \end{figure}

The field of percolation models includes a vast number of various approaches to describe processes in semiconductors 
or  other disordered materials. A general division is given by the description of defects, or whatever is causing the disorder,
either by loss of particles (site percolation) or loss of bonds between sites (bond percolation),    whereby instead of loss one can
observe various bond-strength  or energies at the sites as well \cite{SouLiGr1991,KanOht1999}.\\
In the paper at hand we investigate transport in bond percolation models. The intention here is not the detailed description of any specific 
material but rather the overall description of transport in quantum models of the percolation type.
Therefore we consider in the following a three dimensional cubic lattice with edge length $L$, i.e., the total number of sites 
(or quantum-dimension of the Hamiltonian) is 
$dim\left\lbrace \hat{H}\right\rbrace = \left(\frac{L}{a}\right)^3 = N$, where $a$ denotes the lattice constant, i.e.,
the distance between neighbouring sites.\\
Generally (quantum) one-particle, tight-binding bond percolation may be described by
\begin{equation}
\hat{H} = \sum_{<ij>} t_{ij} \hat{a}^{\dagger}_i \hat{a}_j ~~~, \\
\label{eq:ham1}
\end{equation}
where $<ij>$ denotes the summation over next neighbors and $t_{ij}$ is known as transfer amplitude. This amplitude may given by
\begin{equation}
t_{ij} = \left\lbrace \begin{array}{cl}
t \exp (-2 \pi i \phi_{ij}) & \mathrm{~~~~for ~connected~ bond}\\
0 & \mathrm{~~~~for~ disconnected~ bond}
\end{array} \right.  ~~~. \\
\label{eq:tamp}
\end{equation}
Here, $t$ denotes a parameter which quantifies the hoping strength and $\phi_{ij} $ denotes a parameter
which may
describe interactions with an external field, e.g., magnetic fields; then $\phi_{ij}$ 
is the Peierls phase \citep{KanOht1999}. However, for simplicity and in order to guarantee time reversibility we set $\phi_{ij} =0$.
Moreover the on-site potential is set to zero for all sites, i.e. $t_{ii}=0$. Note that we also set in all calculations $k_B=1$ and $\hbar=1$.\\
The distribution of the bond ``strenghts'' (one for connected, zero for disconnected) is given by 
\begin{equation}
P(t_{ij})=p ~ \delta(t_{ij} =t) +(1-p)~ \delta(t_{ij} = 0)~~.
\label{eq:pdist}
\end{equation}
Fig. (\ref{pic:modperc}) shows an two dimensional sketch of bond percolation $p=0.5$. In the classical case one would certainly say that 
the percolation threshold is just reached, but as
found in  in \cite{SouGr1991,SouLiGr1991,SchuWeiFeh2005} the quantum threshold is (possibly against a naive guess) higher than the 
classical threshold,
i.e., quantum transport would be not possible in the above model, even if it was three-dimensional.

\section{Preliminary investigation of localization}
\label{sec-loca}

\begin{figure}[hbtp]
\includegraphics[scale=0.56]{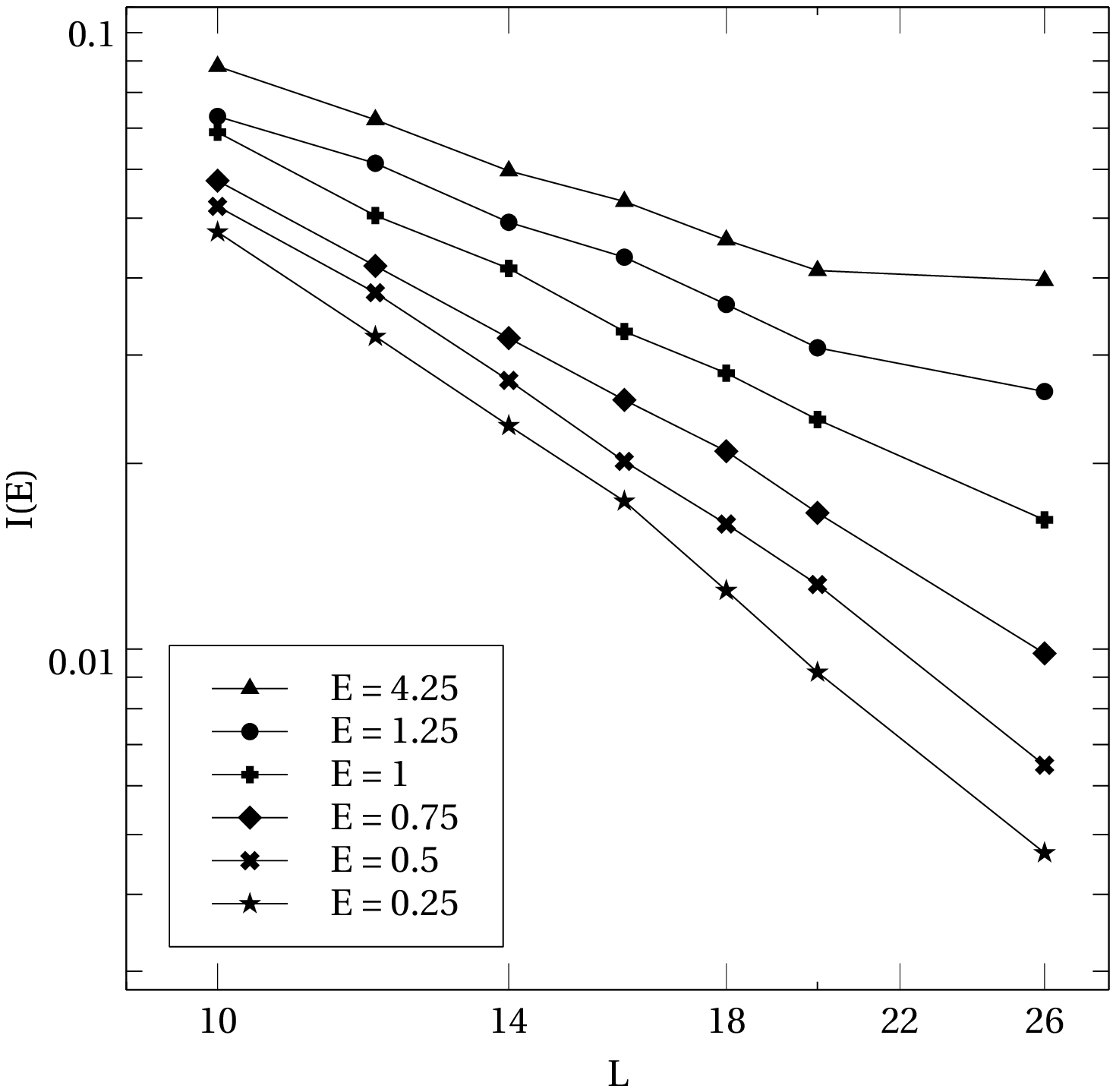}
\caption{Scaling of the IPN with system size $L$  at several energies and percolation ratio p=0.38. Since only at the mobility
edge the scaling is allover linear we locate the mobility edges roughly at  $E=-1$ and $E=1$. Only the last one
$E=1$ is actually displayed above.}
\label{pic:iLin2}
\end{figure}

Below (Sec. \ref{sec-current}) we compute conductivities in the high temperature limit, i.e., all energy  regimes contribute to transport. 
In such a setting an decrease of the conductivity with decreasing $p$ may indicate both, either an increase of resistivity in the delocalized
energy regime, or simply an increase of the fraction of localized energy eigenstates. Since we are primarily interested in the former we 
perform in the following a rough analysis of the fraction of delocalized states for different  $p$'s. Then we concentrate on the regime 
in which the vast majority of the energy eigenstates is delocalized. Generally the precise calculation of mobility edges is a challenge that is 
dealt with using sophisticated methods. \cite{Abrahams2010,priour2012,Mertig1987}. For the purposes at hand, however a rather rough determination of the 
mobility edge suffices. To this end we 
follow the general approach presented in \cite{BrMkos2006}.\\
The investigation at hand is based on the inverse participation number (IPN) 
\begin{equation}
I(E_n) = \sum_i | \psi_i(E_n) | ^4~~~~,
\label{eq:ipn}
\end{equation}
where $\psi_i(E_n) $ denotes the  the $i$-th component of the energy eigenstate
corresponding to $E_n$. As described in \cite{BrMkos2006} a convenient way to find the mobility edge is to plot the IPN at a given energy
against the system size $L$ on a doubly-logarithmic scale. In this representation graphs corresponding to localized states are expected to 
``bend upwards'' while graphs corresponding to extended states are expected to ``bend downwards''. Only the logarithmic IPN's at 
the mobility 
edge are supposed to form a straight line, i.e., the inverse participation number is expected to scale as 
\begin{equation}
I(E_c) \propto L^{-d_2}~~~,
\label{eq:scaleINP}
\end{equation}
where $E_c$ denotes the critical energy (mobility edge) and $d_2$ is referred to as the fractal dimension.
Though the exact determination of fractal dimensions is beyond the scope of this paper, we  note that at $p=0.38$ the fractal dimension in 
our model is approximately given by $d_2 \approx 1.6$ (cf. Fig (\ref{pic:iLin2})).
This value stands in accord with results of a recent work \cite{UjVar2014} where the critical exponent for the
localization length is $d = 1.627\pm0.055$.

\begin{figure}[hbtp]
\includegraphics[scale=0.55]{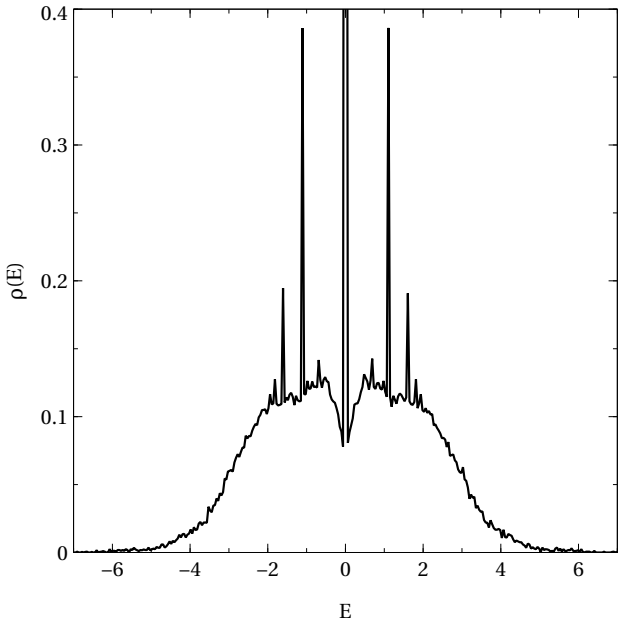}
\caption{The density of states, here for $p=0.45$, is symmetrical w.r.t. the energy $E=0$. There are several distinct
peaks, namely at $E=0$, which correspond to special cluster configurations, cf. \cite{SchuWeiFeh2005}. Also notable
is the \textit{dip} around $E=0$, that only occurs at low $p$.}
\label{pic:dos}
\end{figure}

Fig. (\ref{pic:iLin2}) shows some of the described  scaling graphs for various energies for $p=0.38$, 
all calculated by means of direct numerical diagonalization. This Fig. suggests that 
the mobility edge is around $E\approx 1$ (since $E=1$ appears to correspond to the straightest line). In accord with  an overall symmetry of the 
spectrum w.r.t. energy (see Fig. (\ref{pic:dos})) we find the second mobility 
edge at $E =-1$.
For later purpose it is useful to calculate the density of states (DOS) since it will allow us to estimate the energy
range in which most energy eigenstates are delocalized. To that end we define the portion of delocalized eigenstates
w.r.t. all eigenstates between the above calculated mobility edges which we denote by $\Phi(p)$.

\begin{equation}
\Phi (p) =  \int_{\Delta E} \rho(E) dE~~~~.
\label{eq:phi}
\end{equation}

Before we proceed and introduce the main purpose of this work, we would like to have a closer look at the density of
states for consistency of the given results.\\
One finds that for few impurities, i.e. $p > 0.65$, the density of states is smooth and the graph is well described by a Gaussian function, regardless some peaks which correspond to special cluster configurations within the system, cf. \cite{SchuWeiFeh2005,UjVar2014}.\\
At one hand one finds for low $p$'s that the peaks become more visible, at the other hand one notices a dip around the
energy $E \approx 0$, which becomes more significant for decreasing $p$'s. For visualization of that fact we calculated
the density of states for $p=0.45$. The results are presented in Fig. (\ref{pic:dos}).

\begin{figure}[hbtp]
\includegraphics[scale=0.55]{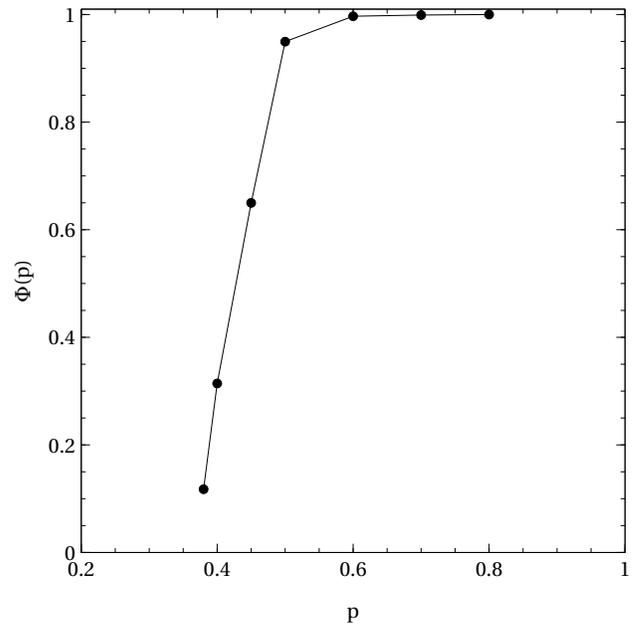}
\caption{Ratio of delocalized eigenstates w.r.t. all energy eigenstates $\Phi(p)$ as calculated by counting the states
between the mobility edges.}
\label{pic:phi}
\end{figure}

The results are displayed in Fig. \ref{pic:phi}. From the latter it is obvious that the regime in which the vast majority of eigenstates are
delocalized is bound from below by $p\approx 0.5$. Furthermore the data appear to be in accord with the value of $p_c \approx 0.31$ 
from the literature for the Anderson transition. 

\section{Current Dynamics and conductivity}
\label{sec-current}

\begin{figure}[hbtp]
\includegraphics[scale=0.54]{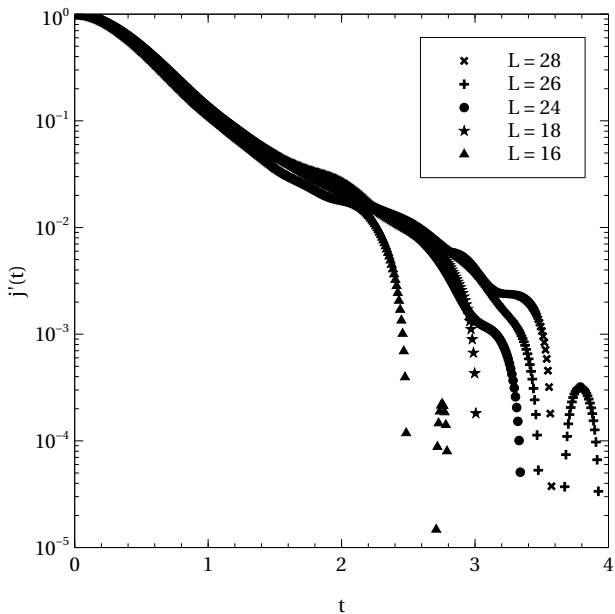}
\caption{Normalized current auto-correlation functions $j'(t)$ for various system sizes $L$. 
The graphs coincide regardless of size in regions where they are substantially different from zero, for, say  $L \geq 18$. Hence data can
reliably be expected to contain negligible finite-size effects at $L = 28$.}
\label{pic:njj}
\end{figure}

As already stated our primary interest in the paper at hand is (other than in many works in the respective literature) not the determination
of the quantum percolation threshold, rather it is quantitative description of transport behavior well above that threshold, i.e., 
in a regime where the 
vast majority of states are extended. 
We aim at finding the dc-conductivity $\sigma_{dc}$ from  linear response theory (Kubo formula) which amounts to the calculation of the 
particle-current autocorrelation function.\\
We restrict ourselves here to the limit of high temperatures and low fillings. Therefore the 
framework of the grand canonical ensemble is used \cite{Jaeckle:EinfTrans,Kubo1957,Zwan1965,Kubo1966} which results in
\begin{equation}
\sigma_{dc} = \sigma(t \rightarrow \infty),~~~\sigma(t) = 
\frac{f}{k_B T} \int^t_0 \frac{1}{V}\mathrm{Tr}\left\lbrace \hat{J}(t')\hat{J}(0) \right\rbrace dt'
\label{eq:sig}
\end{equation}
Here $f$ denotes the filling factor, i.e. the number of particles per site at equilibrium, and $\hat{J}(t)$ denotes the
current operator in the Heisenberg picture. Furthermore $T$ is the temperature and $k_B$ is the
Boltzmann constant. The
volume of the system is denoted by $V = a^3 N$.
\\
In order to employ (\ref{eq:sig}) we need to specify  an adequate current operator. In the context of periodic systems this is often done
using a continuity equation for the site-probabilities \cite{HM-H-Ca2003,GemStM2006,BeFuKluSchee2005}. Since we do not have fully evolved 
periodicity here we follow \cite{khodjaNieGem2013} in starting from a velocity operator instead. The velocity operator (
corresponding to motion in x-direction)  reads:
\begin{equation}
\hat{v} = \frac{i}{\hbar} [\hat{H},\hat{x} ] ~~~~~~~~,
\label{eq:vel}
\end{equation}
where $\hat{x}$ denotes the x-position operator ($x_i$ denotes the x-coordinate of the i-th site)
\begin{equation}
\hat{x} = \sum^N_{i=1} x_i \hat{n}_i ~~~~\hat{n}_i := \hat{a}^\dagger_i \hat{a}_i~~~~x_i = i a.
\label{eq:xpos}
\end{equation}
Thus, the velocity operator $\hat{v}$ reads
\begin{equation}
\hat{v} = \frac{i}{\hbar} \sum^N_{ij} (i - j)\; a \;t_{ji} \hat{a}^\dagger_j \hat{a}_i
\label{eq:velo}
\end{equation}
Note that this expression is at odds with periodic boundaries, i.e., a (short) transition from one edge of the sample through the  
the ``periodic boundary closure''  is, give (\ref{eq:velo}),  equivalent to a (long) transition through the whole sample
in the opposite direction.
Therefore we  modify the expression to ensure that only the ``shortest'' transitions are taken into account. This is achieved by the 
following definition of the current operator:
\begin{equation}
\hat{J} = \sum^N_{ij} J_{ji} \hat{a}^\dagger_j \hat{a}_i
\label{eq:currop}
\end{equation}
\begin{equation}
 J_{ji} = \frac{q}{\hbar} \left\lbrace \begin{array}{ll}
(j-i)\; a\; t_{ij}~~, & |i-j| a < \frac{L}{2}\\
 & \\
\mathrm{sign}(j-i)([L-(j-i)] \;a\; t_{ij})~~, & |i-j|a > \frac{L}{2}\\
\end{array} \right. ~~~\notag
\end{equation}
Here $q$ denotes the electric charge per particle, e.g., elementary charge of a single electron.
In addition to the current operator we introduce one more quantity, namely the normalized current auto-correlation function 
$j'(t)$, which is better suited for the investigation of finite-size effects and convergence behavior than the actual 
current auto-correlation. It is given by
\begin{equation}
j'(t) =\frac{\mathrm{Tr}\left\lbrace \hat{J}(t)\hat{J}(0) \right\rbrace}{\mathrm{Tr}\left\lbrace \hat{J}^2(0) \right\rbrace}~~~.
\label{eq:autocorr}
\end{equation}
Numerical results for (\ref{eq:autocorr}) are displayed in Fig. (\ref{pic:njj}) for various system sizes. (Each curve is the average over 
15 runs for different models featuring the same $p$. However variations with different individual implementations turn out to be small.)
Since the graphs coincide for times 
where they are significant different from zero for, say, $L \geq 18$, it is justified to assume  that at $L=28$ the system is no longer
affected
by finite-size effects. This conclusion is supported by the observation, that the calculation of (\ref{eq:sig}) reveals a deviation of 
the results for a system with $L=26$ compared to one with $L=28$ of approximate $0.9\%$, and continuous decrease of the deviation for 
larger systems.\\
At this point  a comment on numerical techniques is appropriate. Results up to $L=24$ in this paper are always obtained by numerical
matrix diagonalization, whereas all results 
for sizes above this limitation are calculated by means of a algorithm based on ``typicality'' that allows for the determination
of correlation functions on the basis of propagation of single pure states. In the work at hand the pure state propagation is 
performed using a standard Runge-Kutta algorithm. For a full account of this typicality technique and its theoretical background, see 
Refs. \citep{ElsFin2012, Steinigeweg2014a,  Steinigeweg2014b}. We were able to 
treat systems up to $L=34$ ($N\approx 39 000$ ) with this algorithm on standard computing equipment, however as pointed out above,  but 
$L=28$ appears to be sufficient for a reasonable extraction of quantitative results. Nevertheless based on data only from exact diagonalization 
the whole investigation presented here would have been far less conclusive.
\begin{figure}[hbtp]
\includegraphics[scale=0.54]{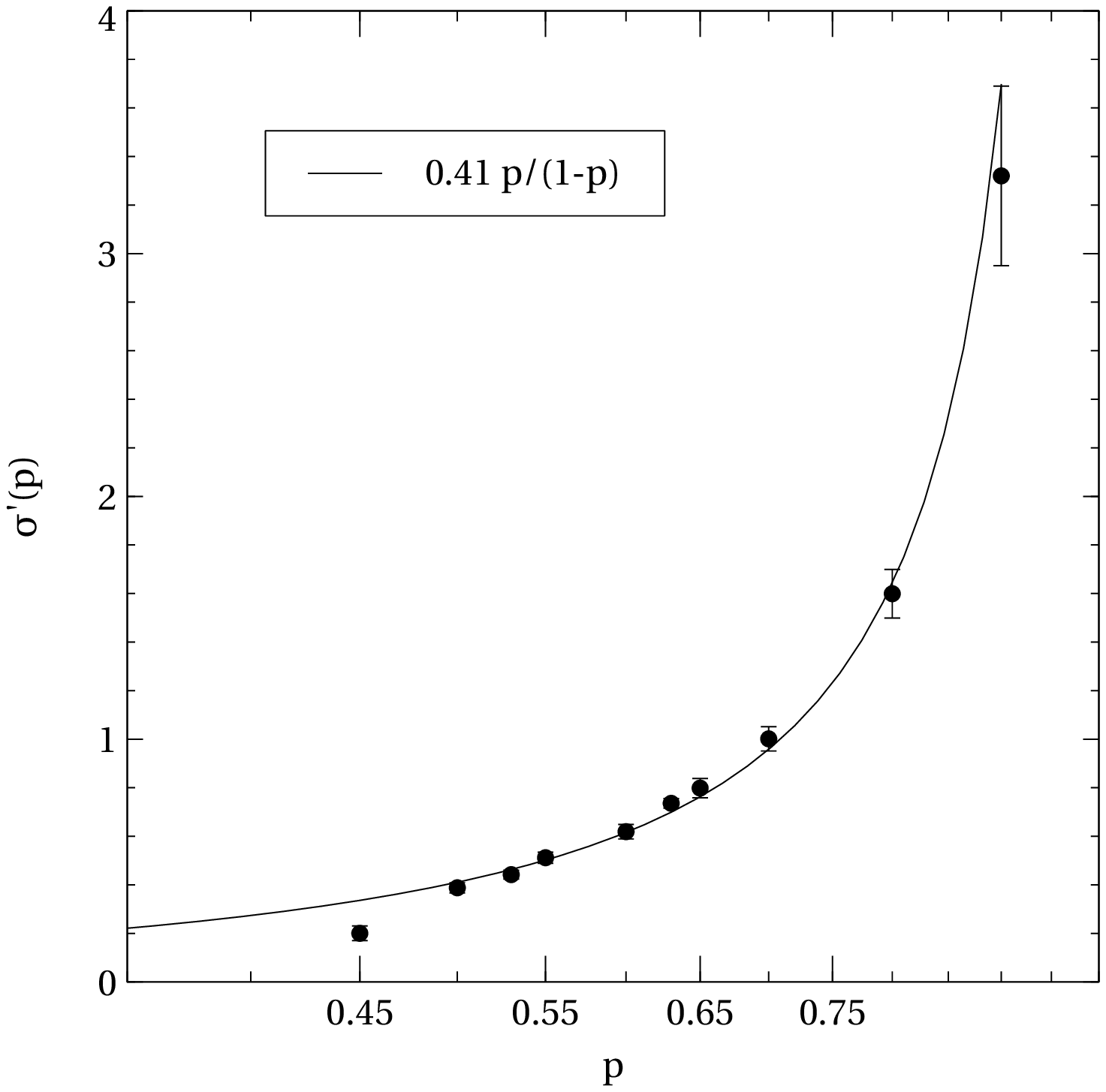}
\caption{Numerically calculated scaled dc-conductivity $\sigma':= Tf^{-1} \sigma_{dc}$ compared to the result of a simple heuristic theory given 
in the text. The agreement is good, deviations appear at and below $p \approx 0.45$. This is due to non-negligible localization. 
Note that all data points carry error bars, however for small $p$ they are barely visible. }
\label{pic:sig}
\end{figure}\vspace*{0.2cm}

The results on conductivity are shown in Fig. (\ref{pic:sig}) where $\sigma'$ relates to the $\sigma_{dc}$ from
(\ref{eq:sig}) as  
\begin{equation}
\sigma_{dc} =   \frac{fq^2 t^2}{k_B T a \hbar [E]}\sigma',
\label{eq:dimensions}
\end{equation}
thus  $\sigma'$ is a dimensionless integrated current autocorrelation function, i.e, $q,  \hbar$ are set to unity and
$[E]$ is the unity according to 
which
energy is measured. Each conductivity represents the average over 15 different percolation models 
featuring the same $p$. The error bars indicate the mean square deviation corresponding to the respective 15 conductivities. As expected, 
the conductivity increases with increasing $p$. A systematic interpretation of this result appears challenging. Nevertheless we want 
to point out the  reasonable agreement of the results displayed in  (\ref{pic:sig}) with results from a simple heuristic reasoning. From simple 
Drude type arguments one expects the conductivity to be proportional to the square of the mean particle velocity $v^2$ and the mean 
collision-free or relaxation time $\tau$, i.e. \cite{AshcroftMermin}, 
\begin{equation}
\sigma \propto v^2\tau.
\label{eq:drude}
\end{equation}
For the  square of the mean particle velocity one may simply take $v^2:=\mathrm{Tr}\left\lbrace \hat{J}^2(0) \right\rbrace/N$. From the current 
operator as given in (\ref{eq:currop}) it is straightforward to see that his quantity must scale as $p$, i.e., e $v^2\propto p$. 
The relaxation time $\tau$ is (by definition) inversely proportional to a scattering  rate $R$. In the percolation model at hand scattering 
(and thus relaxation of the current) is caused by the ``missing connections''. The number of the latter is proportional to $1-p$, hence one obtains 
for the rate $R \propto 1/(1-p)$. Plugging those results into (\ref{eq:drude}) yields 
\begin{equation}
\sigma \propto \frac{p}{1-p}
\label{eq:drudea}
\end{equation}
The solid line in Fig. \ref{pic:sig} shows a fit based on (\ref{eq:drudea}) yielding $\sigma^{'}=0.41 p/(1-p)$. Obviously the agreement is 
rather good  for $p \geq 0.45$. Apparently in this regime the above simple heuristic argument captures the relevant physics, even though
below say $p\ \approx 0.9$ this regime can certainly not be classified as a weak scattering regime. Below  $p\ \approx 0.45$ the fit appears to 
deviate, however, as may be inferred from Fig. \ref{pic:phi}, this is the point at which localization massively sets in. We thus conclude that
the simple theory given in  (\ref{eq:drudea}) holds for $p$ down to the quantum percolation threshold. Furthermore it is clearly noticeable that 
the statistical splay of the results increases with increasing $p$. This however may be readily interpreted as a consequence of the law of large
numbers: the fewer scattering centers there are the larger is the statistical variation  of all quantities that depend on scattering.

\begin{figure}[hbtp]
\includegraphics[scale=0.54]{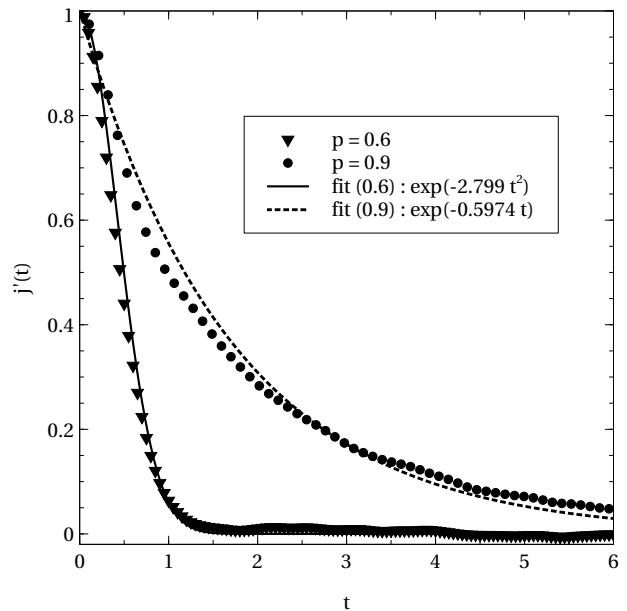}
\caption{Normalized current auto-correlation functions $j'(t)$ at $L = 28$ for $p=0.9$ (few defects) and $p=0.6$ (medium
defects). 
At $p=0.6$ the decay appears to be Gaussian, whereas at $p=0.9$ one finds rough agreement with an exponential decay. The
latter hints in the 
direction of 
Drude-type transport.}
\label{pic:expc}
\end{figure}

Next we consider the specific kind of decay of the current auto-correlation function. As shown in Fig. (\ref{pic:expc}) a 
transition of transport types appears to  occur between $p=0.9$ and $p=0.6$. Decay at $p=0.9$ is compared to a
mono-exponential decay, 
as to be expected from a simple Drude model or a linear Boltzmann equation in relaxation time approximation 
\cite{MahanMany, AshcroftMermin, Jaeckle:EinfTrans}, the agreement is reasonable. At $p=0.65$, however, the decay behavior is much closer to a Gaussian as 
illustrated in  Fig. (\ref{pic:expc}). This transition from exponential to Gaussian decay behavior on the way from weak to strong scattering 
has been observed in various other, similar models before. There it has been explained within the framework of a time-convolutionless projection
operator investigation \cite{Steinigeweg, ca2papers}. If one projects onto the current and performs a perturbative, leading order 
treatment, then exponential decay of the current auto-correlation function results at weak, and Gaussian decay at intermediate strength 
perturbations. From the results displayed in  Fig. (\ref{pic:expc}) it appears evident that the same applies to the model at hand as well.

\section{Einstein relation and mean free path}
\label{sec-einrel}

The diffusion constant of a system is certainly interesting in its own right. However the main purpose of this section is to establish 
the validity of an Einstein relation in oder to arrive at a reasonable definition of a mean free path. The validity of the Einstein relation 
in quantum systems is frequently discussed \cite{SteinwGemWich2009,RoiTes2002}. Here we examine it in its most elementary form, 
namely as the claim of a proportional relation between conductivity and diffusion constant 
\begin{equation}
D(t) = \frac{T}{\epsilon^2} \sigma(t)~~~~,
\label{eq:diffusion}
\end{equation}
where $D(t)$ denotes the (time dependent) diffusion constant, $\epsilon^2$ denotes  the uncertainty (variance) of the transported quantity per 
site at equilibrium. Since the uncertainty for the dc-current equals at low densities the filling factor $f$, cf. \cite{KuboStMeII}, we find
from linear response theory (\ref{eq:sig})
\begin{equation}
D_K(t) =  \int^t_0 \frac{1}{N}\mathrm{Tr}\left\lbrace \hat{J}(t')\hat{J}(0) \right\rbrace dt'~~~~.
\label{eq:diffusionsigma}
\end{equation}
$D_\text{K}(t)$ is to be compared to a direct computation of the diffusion constant in order to check (\ref{eq:diffusion}).
If a diffusion equation holds, the derivation w.r.t. time of the spatial variance of the diffusing quantity equals twice the 
diffusion constant \cite{EinstMolFlu1905}.\\
To directly observe  this spatial variance we define an initial density operator
\begin{equation}
\hat{\rho}(0) =  \frac{1}{Z} \exp ( - \frac{\left( \hat{x}-\frac{L}{2}\right)^2  }{2d}), ~~Z=\mathrm{Tr} \lbrace \exp ( - \frac{\left( \hat{x}-\frac{L}{2}\right)^2  }{2d})\rbrace ~,
\label{eq:rho}
\end{equation}
where $d$ denotes an initial variance, which is here chosen as  $d=0.95$.\\
This implies that the initial site occupation probability is concentrated in a thin slab of a thickness on the 
order of one perpendicular to the x-axis.\\
Based on this $\hat{\rho}(0)$ we calculate the time-depended variance and take the derivative w.r.t. time to obtain a diffusion constant;
here named $D_\text{D}(t)$.
\begin{equation}
D_D(t)= \frac{1}{2}\frac{d}{dt} \mathrm{Tr}\left\lbrace \hat{x}^2(t) \rho(0) \right\rbrace 
\label{eq:d1}
\end{equation}
Note that, since the mean particle position does not drift $\frac{d}{dt} \mathrm{Tr}\left\lbrace \hat{x}(t) \rho(0) \right\rbrace $ remains without
influence.\\

\begin{figure}[hbtp]
\includegraphics[scale=0.55]{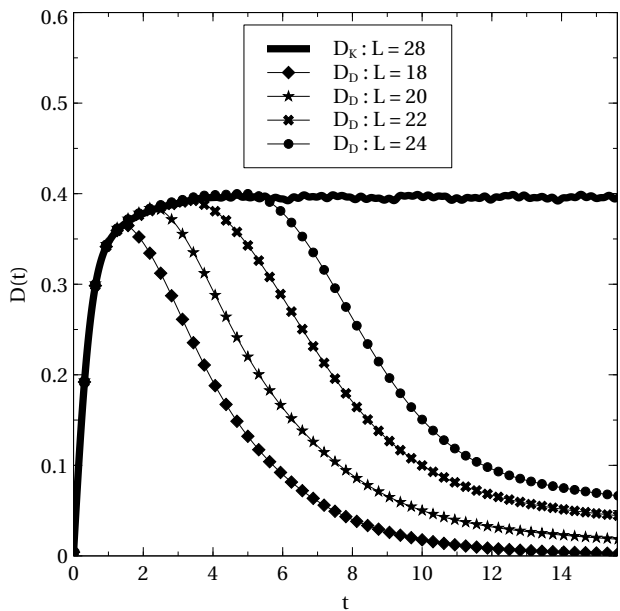}
\caption{Comparison of time-dependent diffusion constants either calculated by (\ref{eq:diffusionsigma}) or (\ref{eq:d1}) 
at $p=0.5$. The calculation according to  (\ref{eq:d1}) obviously suffers from  strong finite-size effects, however, agreement in the limit 
of large sizes is evident. This indicates the validity of the Einstein relation.}
\label{pic:den}
\end{figure}

If the Einstein relation holds, $D_\text{K}(t)$ and $D_\text{D}(t)$ should coincide. Fig. (\ref{pic:den}) shows the comparison of both 
diffusion constants
and reveals that the Einstein relation is apparently fulfilled. Moreover it is obvious that the calculation of the diffusion constant in the 
sense of (\ref{eq:d1}) is strongly influenced by finite-size effects. \\

\begin{figure}[hbtp]
\includegraphics[scale=0.55]{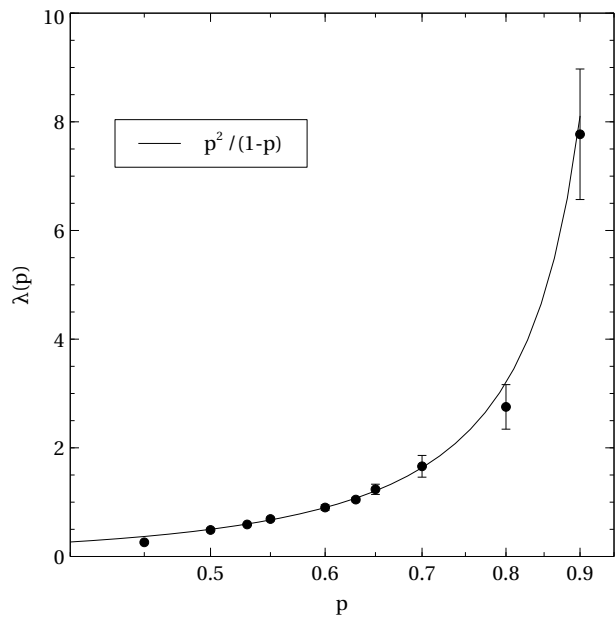}
\caption{Numerically calculated mean free path $\lambda$ compared to the result of a simple heuristic theory given 
in the text. The agreement is good, deviations appear below $p \approx 0.5$. This is due to non-negligible localization. 
Note that all data points carry error bars, however for small $p$ they are barely visible.}
\label{pic:mfp}
\end{figure}

However, the validity of the Einstein relation allows for a reasonable definition of a mean free path $\lambda$ which may be calculated
based on (\ref{eq:diffusionsigma}).  \\
Ballistic transport behavior, as exhibited by initially concentrated, free, non-scattering particles is characterized by a quadratic increase of 
the spatial variance w.r.t. time, i.e., $\left\langle \hat{x}^2\right\rangle   \propto t^2$ or $D \propto t$, 
Since  the increase of the diffusion constant is linear in the beginning, cf. Fig. (\ref{pic:den}), we define the increase of standard deviation 
$\sqrt{ \left\langle \hat{x}^2\right\rangle }$ during this ballistic initial period as the   mean free path $\lambda$. We define the  ballistic 
initial period as the period before $D(t)$ has reached  $90\%$ of its final value. Of course this choice is not imperative, however from looking at 
 Fig. (\ref{pic:den}) it appears reasonable. Fig. (\ref{pic:mfp}) shows the results for the mean free path  $\lambda$

Much like the consideration on conductivity in Sec. \ref{sec-current} we discuss the agreement of a simple, heuristically derived form of $\lambda$ with
the computed data in Fig. (\ref{pic:mfp}) in the following. Consider some ``chain'' of either connections or voids along some crystal-axis. Assume, 
for simplicity that this chain was ordered (which it is in fact not). Assume furthermore that a longer sequence of connections alternates with just
one void. Call the length of the sequence of connections $l$. Then the total ratio $p$ of connections per total number of sites is $p=l/l+1$. Or the
length of the uninterrupted sequence of connections depends on the connection probability as:
\begin{equation}
l(p) =\frac{p}{1-p}
\label{eq:seqprob}
\end{equation}
We may associate $l$ with a free path. I order to find the mean free path we multiply $l$ by $p$ since this is the probability (relative frequency) of 
a ``site'' to sit on a sequence of connections. We thus get:
\begin{equation}
\lambda(p) =\frac{p^2}{1-p}
\label{eq:mfph}
\end{equation}
This expression for the mean free path is represented in Fig. (\ref{pic:mfp}) by the solid line. Given the simplicity of the argument the 
agreement with the computed data is good. Of course such an expression can only be expected to yield reasonable results down to $p=0.5$. However 
from Fig. (\ref{pic:den}) we now that below that localization effects set in anyway. Thus for the fully delocalized regime (\ref{eq:mfph}) appears to 
capture the relevant physics.

\section{Semi-quantitative comparison  of the results to measured conduction data on binary alloys}
\label{sec-semi}

As already pointed out in the Introduction the primary intentions of the paper at hand are of principal nature. However
a short comment 
on the relation of
the results to conductivities of binary alloys should be in order. 
The electronic system of a binary alloy in a mixed crystal phase may be viewed as an
implementation of a percolation model. If the valency of the solute element is very different from that of the host
metal, the on-site potentials 
at the solute sites may be so low (high) that, as a rough approximation, the solute sites
may be 
regarded as being ``frozen out'', i.e., not contributing to the conduction process. Such a picture suggests a site
percolation rather than a bond 
percolation model, but since bond and site percolation are expected to behave more or less comparably we simply ignore this
difference in this consideration. 
The conductivity of weakly or non-interacting fermions at rather low temperatures ($k_BT$ small compared to the
bandwidth) is roughly given by 
\begin{equation}
\sigma_f \approx  \frac{n(E_f)}{N} \int^t_0 \frac{1}{ \mathrm{Tr}\left\lbrace \hat{P}_f\right\rbrace}
\mathrm{Tr}\left\lbrace \hat{J}(t')\hat{J}(0)  \hat{P}_f  \right\rbrace dt'
\label{eq:ferm}
\end{equation}
\cite{Jaeckle:EinfTrans} where $n(E_f)$ is the density of states at the Fermi energy, $N$ is the total number of states
in   the conduction band of the  
one-particle
model and $  \hat{P}_f $ is a projector which projects onto an energy shell (Hilbert space spanned by energy
eigenstates) of width $k_B T$ around the 
Fermi energy. Separating dimensionless quantities from quantities carrying dimensions yields
\begin{equation}
\sigma_f \approx  \frac{n(E_f)q^2 t^2}{N a \hbar [E]}\sigma'_f
\label{eq:fermdim}
\end{equation}
where $\sigma'_f$ is the corresponding dimensionless current auto-correlation function, just like $\sigma'$ in
(\ref{eq:sig}). Since we are only doing an
estimate we replace $\sigma'_f$ by  $\sigma'$ as given in Fig. \ref{pic:sig}, i.e.,  $\sigma'_f \approx 0.41p/1-p$. If the
percentage of solute atoms 
$c[\%]:=100(1-p)$ is low we may approximate $\sigma'_f \approx 41/c$. We intend to compare our results to recently
measured data on 
magnesium alloys, specifically a magnesium-zirconium alloy \cite{Pan}. In order to do so we use the following values in 
(\ref{eq:fermdim}): The bandwidth of 
metallic magnesium is ca. $14eV$ \cite{Chatterjee}, our simple cubic basis model yields a bandwidth of ca.  $14eV$ if
the hopping terms are chosen as
$t=1.2 eV$. (Obviously we use $eV$ as an energy unit). According to \cite{Chatterjee}  we furthermore set the relative
density of states to 
$n(E_f)/N \approx 0.16/eV$. Since our model does not account for any lattice distortions, 
the lattice constant is set to $a= 3 \mathring{A}$ which is about the mean lattice constant of metallic magnesium. And,
naturally, the transported charge 
per particle is the electron charge, i.e., $q=e$. Plugging in all these 
numbers and calculating the specific electrical resistivity $\rho = 1/\sigma_f$ rather than the conductivity itself, we
get
\begin{equation}
\rho \approx c \cdot 1.3 \cdot 10^{-7} (\Omega m) 
\label{eq:rho}
\end{equation}
Of course this cannot be taken as an absolute result since even pure magnesium ($c=0$) has a non-zero resistivity due to
phonons, impurities, etc. But
if one, as suggested by Matthiessen's rule, regards (\ref{eq:rho}) as an expression for the increase of the
resistivity due to the gradual addition 
of a 
a solute,  (\ref{eq:rho}) may be compared to experimental data. Pan et al. report in Ref. \cite{Pan} for a
magnesium-zirconium  alloy a value of 
\begin{equation}
\rho_{measured} =c \cdot 9.311 \cdot 10^{-8} (\Omega m) 
\label{eq:rhomes}
\end{equation}
The atomic volume difference between magnesium and zirconium is rather low, such that few lattice distortions can be
expected. Furthermore the valency
of zirconium ($+4$) is rather high. However, note that our model has simple cubic rather than hexagonal symmetry, we
consider bond rather than site 
percolation, 
the concept of zirconium sites being frozen out is surely not completely correct, we neglect lattice distortions entirely, etc.
Regarding all these limitations
the 
agreement of (\ref{eq:rho}) with  (\ref{eq:rhomes}) within about $30\%$ appears reasonable.

\section{Summary and conclusion}
\label{sec-concl}

We investigated a simple quantum bond percolation model on the basis of an one-particle, tight-binding Hamiltonian. We focus on   
investigation of transport properties in the fully delocalized regime, i.e., a regime in which only a negligible fraction of all energy eigenstates 
is localized. This turns out to be the case at bond probabilities  of  $p \geq 0.5$. The conductivity in this regime has been calculated using linear 
reponse theory (Kubo-formula) and a numerical algorithm based on quantum typicality for the evaluation of the current autocorrelation function. As 
expected the conductivity increases rapidly with increasing $p$ and is found to be in accord with the result of a simple heuristic reasoning involving
mean collision free times and mean particle velocities. The  latter may be defined even though at  $p \geq 0.5$ no true dispersion relations exist.
Furthermore a gradual transition from a current decay that is not in accord with a Drude model to a current decay that is, 
is observed between $p\approx 0.6$ and $p\approx 0.9$. The proportionality of the conductivity and the diffusion constant, i.e., the Einstein relation 
is analyzed numerically and found to hold. This finding allows for a definition of a mean free path. Numerical calculations of this mean free 
path coincide well with results from yet another heuristic consideration based on counting the mean length of uninterrupted sequences of connections
in the lattice. Thus, to conclude, in the regime above $p=0.5$, although being fully quantum and strongly disordered, the dynamics of the model 
appear to be remarkably well described by by purely probabilistic, classical reasoning. Furthermore the result based on the percolation 
model are in reasonable agreement with measured data on binary magnesium alloys.

\bibliography{latbi2}

\end{document}